\documentstyle[prl,aps,twocolumn,floats,psfig]{revtex}

\newcommand{\bleq}{\ifpreprintsty
                   \else
                   \end{multicols}\vspace*{-3.5ex}{\tiny
                   \noindent\begin{tabular}[t]{c|}
                   \parbox{0.493\hsize}{~} \\ \hline \end{tabular}}
                   \fi}
\newcommand{\eleq}{\ifpreprintsty
                   \else
                   {\tiny\hspace*{\fill}\begin{tabular}[t]{|c}\hline
                    \parbox{0.49\hsize}{~} \\
                    \end{tabular}}\vspace*{-2.5ex}\begin{multicols}{2}
                    \fi}
\newcommand{\bcols}{\ifpreprintsty\else\begin{multicols}{2}\fi}
\newcommand{\ecols}{\ifpreprintsty\else\end{multicols}\fi}

\def\mathcal{\cal}
\def \be{\begin{equation}}
\def \ee{\end{equation}}
\def \bmlett{\begin{mathletters}}
\def \emlett{\end{mathletters}}

\def \HH{{\mathcal H}}
\def \NN{{\mathcal N}}

\def \ua{\uparrow}
\def \da{\downarrow}
\def \ra{\rightarrow}

\def \tmeas{ \tau_{\rm meas} }
\def \Gmix{ \Gamma_{\rm mix} }

\begin{document}

\bibliographystyle{simpl1}


\title{Resonant Cooper Pair Tunneling: Quantum Noise and
Measurement Characteristics}

\author{A. A. Clerk, S. M. Girvin, A. K. Nguyen, and A. D. Stone}
\address{
Department of Applied Physics and Physics, Yale University, New
Haven CT, 06511, USA\\
March 27, 2002
\medskip ~~\\ \parbox{14cm}{\rm
We study the quantum charge noise and measurement properties of
the {\it double} Cooper pair resonance point in a superconducting
single-electron transistor (SSET) coupled to a Josephson charge
qubit. Using a density matrix approach for the coupled system, we
obtain a full description of the measurement back-action; for weak
coupling, this is used to extract the quantum charge noise. Unlike
the case of a non-superconducting SET, the back-action here can
induce population inversion in the qubit. We find that the Cooper
pair resonance process allows for a much better measurement than a
similar non-superconducting SET, and can approach the quantum
limit of efficiency.
}} \maketitle

Among the many open issues related to solid state quantum
computation, the question of how best to {\it measure} a solid
state qubit remains a particularly interesting one.  In the case
where the qubit is a Cooper pair box (i.e. a Josephson-junction
single charge box), the standard choice for a read-out device is
the single-electron transistor (SET)
\cite{RobSET,Korotkov,DevoretNature,MakhlinRMP,Aassime,Wendin}. An
alternate and potentially more powerful approach is to use a {\it
superconducting} single electron transistor (SSET) biased at a
point where the cyclic resonant tunneling of Cooper pairs
dominates transport \cite{Fulton,Averin,Maasen,NakamuraJQP,Pohlen}
\cite{Zorin}. Such processes, known as Josephson quasiparticle
(JQP) resonances, would appear to be an attractive choice for use
in a measurement as their resonance structure implies an extremely
high sensitivity. However, precisely because of their large gain,
these processes may be expected to strongly alter the state of the
qubit in a measurement.  To assess the balance between these two
opposing tendencies, a close examination of the physics of JQP
tunneling is required. Note that the SSET-qubit system is more
relevant to experiment than a setup having a SET, as fabrication
typically results in the entire system being made from a single
metal.

\begin{figure}
\centerline{\psfig{figure=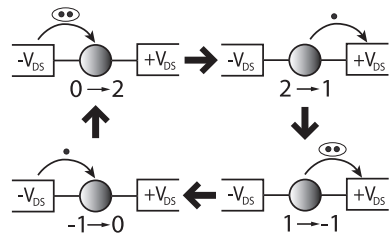,width=8cm}}
\caption{Schematic showing the four steps of the double Josephson
quasiparticle process which can occur in a superconducting
single-electron transistor.  Circles represent the central island
of the SSET, while the rectangles are the electrodes.  Numbers
indicate the charge of the SSET island.} \vspace{-0.5 cm}
\label{EyeSchem}
\end{figure}

In this paper, we focus on a {\it double} JQP process (DJQP) (see
Fig.~\ref{EyeSchem}), which occurs at a lower SSET source-drain
voltage than single JQP processes, and which has been used in a
recent experiment \cite{Konrad}.  We assess the potential of DJQP
to act as a one-shot measurement of the state of a Cooper pair box
qubit. This involves characterizing both $\tmeas$, the time needed
to discriminate the two qubit states in the measurement, and the
back-action of the measurement on the qubit, which is described by
a mixing rate $\Gmix$ and a dephasing rate $\tau_{\varphi}$. These
quantities are intimately related to the noise properties of the
SSET, which are of fundamental interest in themselves, given the
novel nature of the DJQP process. $\tmeas$ is determined largely
by the shot noise of the process, while $\Gmix$ and
$\tau_{\varphi}$ are related to the associated charge noise on the
SSET island. While the shot noise of a {\it single} JQP process
has been analyzed recently \cite{Choi}, the quantum charge noise
has not been addressed. It is of particular interest, as the
experiment of Ref. \cite{Konrad} uses the ability of a qubit to
act as a spectrum analyzer of this quantum charge noise.

To describe the measurement process in our system, we employ a
density matrix description of the {\it fully coupled} SSET plus
qubit system; this is similar to the approach taken by Makhlin
{\it et al.} \cite{MakhlinRMP} for a SET, but extended to deal
with Josephson tunneling.  This approach is not limited by a
requirement of weak-coupling, as are standard approaches which
perturbatively link $\Gmix$ to the transistor charge noise
\cite{Aassime,Wendin}; nonetheless, in the limit of weak-coupling
the present method can be used to calculate the quantum charge
noise of the SSET. We find that the quantum (i.e. asymmetric in
frequency) nature of the noise is particularly pronounced for the
DJQP feature, leading to regimes where the SSET can strongly relax
the qubit.  Moreover, due to the resonant nature of Cooper pair
tunneling, there exist regimes where the SSET can cause a
pronounced {\it population inversion} in the Cooper pair box,
something that is impossible using a SET. For typical device
parameters, we find that a far better single-shot measurement is
possible using the DJQP process than with a comparable SET.

{\it Model--} The Hamiltonian of the coupled qubit plus SSET
system is is written as $\HH = \HH_{\rm S} + \HH_{\rm Q} +
\HH_{\rm int}$. The qubit itself (or ``box"), described by
$\HH_{\rm Q}$, consists of a superconducting metal island in the
Coulomb blockade regime where only two charge states are relevant.
These can be regarded as the $\sigma_z$ eigenstates of a
fictitious spin $1/2$. The island is attached via a tunnel
junction to a bulk superconducting electrode, leading to the form
\begin{equation}
    \HH_{Q} = -\frac{1}{2} \left[
        \left( 4 E_{CQ} (1 - \NN_{Q}) \right) \sigma_{z}
        + E_{JQ} \sigma_x \right].
\end{equation}
{\noindent}where $E_{CQ}$ is the charging energy of the box,
$E_{JQ}$ is the Josephson coupling energy of the box, and $\NN_Q$
is the dimensionless gate voltage applied to the box.  The SSET
consists of a superconducting, Coulomb-blockaded island which is
attached via tunnel junctions to two superconducting electrodes
(Fig.~\ref{EyeSchem}). The SSET Hamiltonian $\HH_{S} = H_{K} +
H_{C} + H_{V} + H_{T}$ has a term $H_K$ describing the kinetic
energy of source, drain and central island electrons,  a term
$H_V$ which describes the work done by the voltage sources, and a
tunneling term $H_T$. The charging term is $H_C = E_{CS} (n_{S} -
\NN_S)^2$, where $E_{CS}$ is the SSET charging energy, $n_{S}$ is
the number of electrons on the central island, and $\NN_S$ is the
dimensionless gate voltage applied to the island. Finally, the
qubit is capacitively coupled to the SSET: $\HH_{\rm int} = 2
E_{CQ} \frac{ C_{C} }{C_{\Sigma}} \sigma_z n_S \equiv E_{\rm int}
\sigma_z n_S$. Here $C_C$ is the cross-capacitance between the box
and the central island of the SSET, and $C_{\Sigma}$ is the total
capacitance of the SSET island. Note that we neglect the coupling
of the qubit to its environment, as we are interested here in the
intrinsic effect of the SSET on the qubit \cite{EnvNote}.  We also
assume a SSET with identical tunnel junctions, whose dimensionless
conductance $g$ satisfies $g/(2 \pi) \ll 1$.

The DJQP process occurs when the SSET gate voltage $\NN_S$ and
drain-source voltage $2 V_{DS}$ are such that two Cooper-pair
tunneling transitions (one in each junction) are resonant.  We
label these transitions as $n_s = 0 \ra 2$ (left junction) and
$n_s = 1 \ra -1$ (right junction) (see Fig.~\ref{EyeSchem}).
Resonance thus requires $eV_{DS} = E_{CS}$ and $\NN_S = 1/2$. In
addition, $E_{CS} / \Delta_S$ (where $\Delta_S$ is the
superconducting gap of the SSET) must be chosen so that the
quasiparticle transitions linking the two Cooper pair resonances
are energetically allowed (i.e. $n_S = 2 \ra 1$ and $n_S = -1 \ra
0$), whereas transitions which end the cycle (i.e. $n_S = 0 \ra
1$) are not. We take $E_{CS} = \Delta_S$ to satisfy these
conditions; this corresponds to the experiment of Ref.
\cite{Konrad}. The two quasiparticle transitions which occur in
the DJQP are characterized by a rate $\Gamma$, which is given by
the usual expression for quasiparticle tunneling between two
superconductors \cite{Tinkham}.  The effective Cooper pair
tunneling rate $\gamma_J$ emerging from our description (i.e.
Eq.~(\ref{RhoEvolution}) below) is given by \cite{Averin}:
\begin{equation}
\label{CPRate}
    \gamma_J(\delta) = \frac{ E_{JS}^2 \Gamma }{4 \left(
        \delta^2 + (\Gamma/2)^2 \right)}
\end{equation}
{\noindent}Here, $\delta$ is the energy difference between the two
charge states involved in tunneling, $E_{JS}$ is the Josephson
energy of the SSET, and we set $\hbar = 1$.

{\it Calculation Approach-- }We consider the reduced density
matrix $\rho$ of the qubit plus SSET system obtained by tracing
out the SSET fermionic degrees of freedom.  The evolution of
$\rho$ is calculated perturbatively in the tunneling Hamiltonian
$H_T$, keeping only the lowest order terms; this corresponds to
the neglect of co-tunneling processes, which is valid for small
$g$ and near the DJQP resonance.  Using an interaction
representation where $\HH_T$ is viewed as a perturbation, the
equation of motion of $\rho$ takes the standard form:
\begin{equation}
    \label{RhoEvolution}
    \frac{d}{dt} \rho(t) =
        - \int_{-\infty}^{t} dt' \langle
            \left[
                \HH_T(t), \left[
                    \HH_T(t'),  \rho(t') \otimes \rho_{F}
                    \right]
                    \right]
                        \rangle
\end{equation}
The angular brackets denote the trace over SSET fermion degrees of
freedom; as we work at zero temperature, $\rho_{F}$ is the density
matrix corresponding to ground state of these degrees of freedom
in the absence of tunneling. In the diagrammatic language of Ref.
\cite{SchonPRB}, Eq.~(\ref{RhoEvolution}) is equivalent to keeping
all $\HH_T^2$ terms in the self-energy of the Keldysh propagator
governing the evolution of $\rho$.  Note that the correlators in
Eq.~(\ref{RhoEvolution}) describe both quasiparticle tunneling and
Josephson tunneling in the SSET.

To make further progress, we treat the Josephson coupling emerging
from Eq.~(\ref{RhoEvolution}) as energy-independent and given by
the Ambegaokar-Baratoff value $E_{JS} = g \Delta_S / 8$. We also
use the smallness of $g$ to neglect logarithmic renormalization
terms, as was done in Ref. \cite{MakhlinRMP}.  One can then
immediately solve for the time-independent solution of
Eq.~(\ref{RhoEvolution}), which describes the quasi-equilibrium
state achieved by the system after all mixing and dephasing of the
qubit by the SSET has occurred.
To describe the dynamics of mixing (i.e. the relaxation of the
qubit state populations to their stationary value), we also
calculate the corresponding eigenmode of Eq.~(\ref{RhoEvolution}).
A Markov approximation is made which involves replacing $\rho(t')$
by $\rho(t)$ on the RHS of Eq.~(\ref{RhoEvolution}); for the
mixing mode, this should be done in the Schr\"odinger picture
\cite{Me}. This approximation is justified as long as the time
dependence of $\rho$ in the mixing mode is weak compared to
typical frequencies appearing in the correlators of
Eq.~(\ref{RhoEvolution}), requiring here that $\Gmix < \Gamma,
E_{CS}$ and $E_{JS} < E_{CS}$ \cite{Me}.

\begin{figure}
\centerline{\psfig{figure=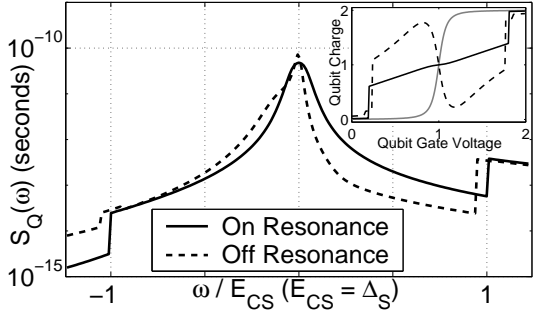,width=8.0cm}}
\caption{Quantum charge noise associated with the DJQP process.
The solid curve corresponds to $\NN_S$,$V_{DS}$ tuned to the
center of the DJQP resonance; the dashed curve corresponds to
moving $e V_{DS}$ away from resonance by $+\Gamma / 4$.  We take
$g = 0.5$ and $\Delta_S \simeq E_{CS} = 0.25 m eV$ in the SSET,
corresponding to the device of Ref.~13; this gives $E_{JS} / (h
\Gamma) \simeq 0.04$. Inset: average qubit charge after mixing has
occurred for weak coupling ($E_{\rm int}/E_{JQ} = 0.01$), as a
function of qubit gate voltage $\NN_Q$; see text for details. As
in Ref.~13, we have $E_{CQ} \simeq 77 \mu eV$ and $E_{JQ} \simeq
27 \mu eV$. The frequency range probed by tuning $\NN_Q$ matches
the range of the main plot; the sharp steps in the average charge
occur at $\Omega(\NN_Q) \simeq E_{CS}$.} \vspace{-0.5 cm}
\label{WeakPlot}
\end{figure}

{\it Back-Action--}  We focus here primarily on the mixing effect
of the measurement back-action; dephasing will be discussed more
extensively in Ref. \cite{Me}.  The mixing rate $\Gamma_{\rm mix}
= \Gamma_{\rm rel} + \Gamma_{\rm exc}$ is set by the rates at
which the measurement relaxes and excites the qubit.  Let $\Omega$
denote the $\NN_Q$-dependent energy difference between the two
qubit states.  For weak coupling ($E_{\rm int} \ll \Omega$),
Fermi's Golden rule relates $\Gamma_{\rm rel}$ and $\Gamma_{\rm
exc}$ to the quantum charge noise of the SSET island $S_Q(\omega)
= \int d t \mbox{ } e^{-i \omega t} \langle n_S(t) n_S(0)
\rangle$:
\begin{equation}
\label{GoldenRule}
    \Gamma_{\rm rel/exc} = E_{\rm int}^2
        \left(\frac{E_{JQ}}{\Omega}\right)^2
        S_{Q}(\pm \Omega).
\end{equation}
In our approach, these rates may be directly obtained by using the
stationary solution (which gives the post-mixing occupancies of
the box eigenstates) and the mixing eigenvalue of
Eq.~(\ref{RhoEvolution}). In the limit of weak coupling, one can
then use Eq.~(\ref{GoldenRule}) to extract $S_Q(\Omega)$.  Our
method for calculating the quantum noise, which uses the qubit as
a spectrum analyzer, is physically intuitive and no more difficult
to implement than standard approaches \cite{Wendin}; in addition,
we are able to calculate $\Gamma_{\rm rel}$ and $\Gamma_{\rm exc}$
when the coupling is not weak, and Eq. (\ref{GoldenRule}) fails.

Fig.~\ref{WeakPlot} displays the quantum charge noise obtained at
zero temperature, using SSET parameters which correspond to Ref.
\cite{Konrad}.  The solid curve in Fig.~\ref{WeakPlot} is for the
center of the DJQP resonance-- $\NN_S = 1/2$, $e V_{DS} = E_{CS}$.
Note the sudden asymmetry that develops between absorption (i.e.
$S_Q(+|\omega|)$) and emission (i.e $S_Q(-|\omega|)$) when
$|\omega|$ increases beyond $E_{CS}$. These sudden jumps
correspond to the opening and closing of transport channels in the
SSET, and their sharpness (which is absent for similar processes
in a SET \cite{Wendin}) is a direct consequence of the singularity
in the quasiparticle density of states.  For example, as $\omega$
rises past $E_{CS}$, quasiparticle transitions which are normally
forbidden in the DJQP cycle (i.e. $n_S = 0 \ra 1$) suddenly become
energetically allowed {\it if} they absorb energy from the qubit,
causing a sudden increase in $S_Q(\omega)$.

The effect of the SSET quantum charge noise on the qubit is shown
in the inset of Fig.~\ref{WeakPlot}, where the average qubit
charge $\langle N_B \rangle \equiv 1 + \langle \sigma_z \rangle$
for $t \gg \tau_{mix}$ is shown as a function of $\NN_{Q}$.
Changing $\NN_{Q}$ tunes the qubit splitting frequency $\Omega$,
allowing one to effectively probe the frequency dependence of the
noise. The solid black curve corresponds to being at the center of
the DJQP feature, while the grey curve corresponds to the
unperturbed qubit ground state. The sharp features in the quantum
noise manifest themselves in $\langle N_B \rangle$, a quantity
which is directly accessible in experiment.

Even more interesting are features emerging in the low frequency
quantum noise ($|\omega| \ll E_{CS}$) when one tunes $\NN_S$ or
$V_{DS}$ slightly off the DJQP resonance center. Unlike the case
of a SET, where asymmetries in the noise are weak for these
frequencies, there are strong features here that result from the
resonant nature of Cooper pair tunneling.  By treating the mixing
terms in Eq.~(\ref{RhoEvolution}) perturbatively, simple analytic
expressions can be obtained for the quantum noise in this regime
when $E_{JS} < \Gamma$ (in Ref. \cite{Konrad}, $E_{JS}/(h \Gamma)
\simeq 0.04$). If one moves away from the DJQP center by tuning
only $V_{DS}$ (i.e. $\NN_S = 1/2$, $ e V_{DS} = E_{CS} +
\delta_V/2$), we find $(|\omega| < E_{CS})$:
\begin{equation}
\label{ResMixing}
    S_Q(\omega) =
        \gamma_J(\delta_V)
        \frac{
            \left[\gamma_J(\delta_V + \omega) / \gamma_J(\delta_V
            - \omega)\right] }
            {\left[4 \gamma_J(\delta_V + \omega) \gamma_J(\delta_V -
            \omega)\right] + \omega^2}
\end{equation}
In the limit where $\omega$ is much smaller than the width
$\Gamma/2$ of the Cooper pair resonance, Eq.~(\ref{ResMixing})
simply corresponds to classical telegraph noise (the SSET only
spends appreciable time in the states $n_S=0$ and $n_S=1$).
However, for finite $\delta_V$ and $\omega$, Eq.~(\ref{ResMixing})
indicates that the noise develops a pronounced asymmetry, even
though $|\omega| \ll E_C$.  In particular, if $\delta_V > 0$, one
has $S_Q(-|\omega|)
> S_Q(+|\omega|)$, implying that {\it emission by the SSET exceeds
absorption}. This behavior is shown by the dashed curves in
Fig.~\ref{WeakPlot}, which correspond to $\NN_S= 1/2$, $\delta_V =
+ \Gamma/4$.  This effect is a direct consequence of the resonant
nature of Cooper pair tunneling-- by emitting energy, {\it both}
Cooper pair tunneling processes in the DJQP cycle become more
resonant, while absorbing energy pushes them even further from
resonance. The net result is a population inversion in the qubit
at zero temperature, which in turn leads to a striking,
non-monotonic dependence of qubit charge on $\NN_Q$ (this is shown
by the dashed curve in the inset of Fig.~\ref{WeakPlot})
\cite{EnvNote}. Note that if one moves away from the center of the
DJQP resonance by only changing the gate voltage $\NN_S$, no
asymmetry in the noise results, as now emission (or absorption)
moves one of the Cooper-pair transitions in the DJQP process
further {\it towards} resonance, while it moves the other
transition further {\it away} from resonance. Nonetheless, the
noise in this case still has a non-monotonic dependence on
frequency. Letting $\delta_V = 0$ and $\delta_{\NN} = 4 E_{CS}
(\NN_S - 1/2)$, we have for $E_{JS} < \Gamma$:
\begin{equation}
\label{DullResMixing}
    S_Q(\omega) =
        \gamma_J(\delta_{\NN})
        \frac{1 +
        \frac{ \left(8 \delta_{\NN} \omega  \right)^2 }
            {E_{JS}^4/2}
            \gamma_J(\delta_{\NN} - \omega) \gamma_J(\delta_{\NN} +
            \omega) }
            {\left[4 \gamma_J(\delta_{\NN} + \omega) \gamma_J(\delta_{\NN} -
            \omega)\right] + \omega^2}
\end{equation}

{\it Measurement Rate-- }To determine the measurement time
$\tmeas$, we extend our density matrix description to also include
$m$, the number of electrons that have tunnelled through the left
SSET junction \cite{MakhlinRMP,Choi} \cite{CurrNote}.  We are thus
able to calculate the distribution of tunnelled electrons
$P(m,t|i)$, where $i=\ua,\da$ denotes the initial state of the
qubit. $\tmeas$ is defined as the minimum time needed before the
two distributions $P(m,t|\ua)$ and $P(m,t|\da)$ are statistically
distinguishable \cite{MakhlinRMP}:
\begin{equation}    \label{TMDefn}
    \frac{1}{\tmeas} = \left(
        \frac{I_{\ua} - I_{\da}}
        {\sqrt{2 f_{\ua} I_{\ua}} + \sqrt{2 f{\da} I_{\da}} }
        \right)^2,
\end{equation}
{\noindent}Here, $I_{\ua}$ and $I_{\da}$ are the average SSET
currents associated with the two qubit states, and $f_{\ua}$ and
$f_{\da}$ are the associated Fano factors which govern the
zero-frequency shot noise in the current (i.e. the distribution
$P(m,t|i)$ is roughly Gaussian with a mean $(I_i/e) t $ and
standard deviation $\sqrt{f_i (I_i/e)t}$).  In the absence of the
qubit, analysis of the density matrix equations for the SSET
reveals:
\begin{equation}
\label{DJQPFano}
    f(\delta) = \frac{3}{2} \left[
        1 - \frac{1}{2}
            \frac{ E_{JS}^2 \left(3 (\Gamma/2)^2 -
                \delta^2 \right) }
                {\left( [ \Gamma / 2 ]^2 + \delta^2 + E_{JS}^2/2
                \right)^2}
            \right],
\end{equation}
where we take $e V_{DS} = E_{CS}$, $\delta = \delta_{\NN} = 4
E_{CS} (\NN_S - 1/2)$. Eq.~(\ref{DJQPFano}) indicates that the
effective charge of the carriers in the DJQP process is $3e/2$ in
the limit where $\Gamma \gg E_{JS}$.  In this limit, Cooper-pair
tunneling is the rate-limiting step in the cycle; electrons
effectively tunnel in clumps of $e$ or $2 e$, leading to an
average charge of $3e/2$. A similar argument holds in the opposite
regime $\Gamma \ll E_{JS}$.

\begin{figure}
\centerline{\psfig{figure=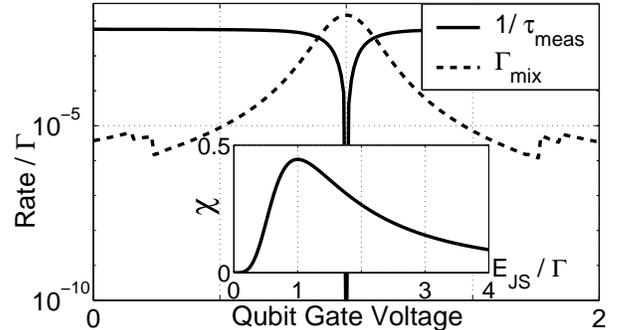,width=8 cm}}
\caption{$1/\tmeas$, $\Gamma_{\rm rel}$, and $\Gamma_{\rm exc}$
vs. qubit gate voltage $\NN_Q$ for a strongly coupled system,
where $E_{\rm int} / E_{JQ} \simeq 0.3$ (i.e. $C_C / C_{\Sigma} =
0.05$). A good measurement is possible for a wide range of gate
voltages.  Inset: Heisenberg efficiency $\chi = \tau_{\varphi} /
\tau_{\rm meas}$ at weak coupling, as a function of
$E_{JS}/\Gamma$. } \label{StrongPlot} \vspace{-0.5 cm}
\end{figure}

We consider $\tmeas$ in the limit of weak coupling ($E_{\rm int}
\ll \Omega$) and weak mixing ($E_{JQ} \ll \Omega$). In this limit,
the two qubit states simply provide an effective shift in the SSET
gate voltage.  Taking $\delta_V = 0$ and $\delta_\NN = \Gamma/ 2$
for near optimal gain, and using Eqs.
(\ref{DullResMixing}-\ref{DJQPFano}), we find that the intrinsic
signal-to-noise ratio $\left( \tmeas \Gmix \right)^{-1/2}$ of the
measurement, in the relevant regime $E_{JS} < \Gamma$, is given
by:
\begin{equation}
\label{SSETSN}
    (S/N)_{DJQP} =  \sqrt{ \frac{4}{3} } \left|
                \cot \theta
                \right|
                \frac{\Omega}{\Gamma/2}.
\end{equation}
{\noindent}Here, $\cot \theta \equiv 4 E_{CQ} (1 - \NN_{Q}) /
E_{JQ}$, and we take $\gamma_J(0) \ll \Omega < E_{CS}$. If a SET
in the sequential-tunneling regime is used for the qubit
measurement, it was found in Refs. \cite{DevoretNature,MakhlinRMP}
that the optimal $S/N$ is given by ($\Omega < E_{CS}$):
\begin{equation}
\label{SETSN}
    (S/N)_{SET} =   \lambda \left|
                    \cot \theta
                \right| \sqrt{
                    \left(\frac{\Omega}{e V_{DS}} \right)^2 +
                    \frac{g^2}{\pi^2}  },
\end{equation}
{\noindent}where $\lambda$ is of order unity. As the quasiparticle
transition rate $\Gamma \sim \frac{g}{2 \pi} e V_{DS}$, we see
that the $S/N$ achieved using DJQP is parametrically larger (in $2
\pi/g \gg 1$) than that obtained for the SET. This enhancement
results largely from the narrow width of the DJQP feature-- the
energy scale over which the current changes (and thus the gain) is
set by $\Gamma$ rather than $V_{DS}$.  The gain and $S/N$ ratio of
the SET could be improved by working in the co-tunneling regime;
however, this would result in a much larger $\tau_{meas}$
($\tau_{meas} \propto g^{-2}$), making one more susceptible to
unwanted environmental effects.  In contrast, the DJQP feature has
both a large gain {\it and} a short $\tau_{meas}$ (i.e.
$\tau_{meas} \propto 1/g$).  Shown in Fig.~\ref{StrongPlot} as a
function of $\NN_{Q}$ are $\tmeas, \Gamma_{\rm rel}$ and
$\Gamma_{\rm exc}$ for a strongly coupled device ($C_C /
C_{\Sigma} = 0.05$), with all other parameters corresponding to
Ref. \cite{Konrad}. We have taken $\delta_V = 0$ and $\delta_\NN =
\Gamma/2$ for optimal gain. Fig.~\ref{StrongPlot} confirms that an
excellent measurement is indeed possible, with $(S/N)^2 > 100$.
Note that the strong coupling splits the sharp features occurring
in the noise (see Fig~\ref{WeakPlot}, solid curve).

We have also studied the ``Heisenberg efficiency" $\chi =
\tau_{\varphi} / \tau_{\rm meas}$ of measurement using DJQP for a
weak coupling ($E_{\rm int} \ll E_{JS},\Gamma$) and $\Omega <
E_{CS}$, where $\tau_{\varphi}$ is the measurement-induced
dephasing time \cite{Me}. Unlike an SET in the sequential
tunneling regime, where $\chi \propto g^2$ is always much less
than the quantum limit $\chi = 1$ \cite{DevoretNature,MakhlinRMP},
here $\chi$ is controlled by the ratio $E_{JS} / \Gamma$.  As
shown in the inset of Fig. \ref{StrongPlot}, by tuning this ratio,
$\chi$ {\it can be made to approach the quantum limit}. Here, for
each value of $E_{JS} / \Gamma$ we have set $V_{DS}$ and $\NN_S$
to optimize the gain.  Measurement using DJQP is able to reach a
high efficiency when $E_{JS} \simeq \Gamma$ both because of the
symmetry of the process, and because of the coherent nature of
Josephson tunneling \cite{Me}. Clearly, the DJQP process allows
for a far superior measurement of a Cooper pair box qubit than a
SET.

We thank M. Devoret, K. Lehnert and R. Schoelkopf for useful
discussions. This work was supported by the NSF under Grants No.
DMR-0084501 and DMR-0196503, and by the Army Research Office under
Grant No. ARO-43387-PH-QC.



\begin{references}
\bibitem{EnvNote}
    To observe the population inversion predicted here,
    relaxation due to the environment must be weak
    compared to mixing from the measurement.  For the devices of
    Ref. \cite{Konrad}, environmental relaxation rates are $\sim 10^6 Hz$;
    a coupling strength $C_C/C_{\Sigma} \simeq 0.1$ is then
    sufficient.
\bibitem{CurrNote}
    As current is conserved, we could equally as well have chosen
    the right junction.
\end{references}
\end{document}